\documentclass{aa}     
\usepackage{psfig}
\usepackage{graphics}

\def\rosat{{\it ROSAT\/\ }}

\begin{document}

\font\myfontsmall=cmr17 at 8.0pt

   \thesaurus{
             }

   \title{1ES 1927+654: Persistent and rapid  X-ray variability in an AGN with low intrinsic neutral X-ray absorption
   and  narrow optical emission lines}
  \authorrunning{Boller et al.}
   \titlerunning{1ES 1927+654 - rapid X-ray variability and narrow optical lines}

   \author{Th. Boller$^1$, W. Voges$^1$, M. Dennefeld$^2$, I. Lehmann$^1$, P. Predehl$^1$, V.  Burwitz$^1$,
 E. Perlman$^{3,4}$, L. Gallo$^1$,  I.E. Papadakis$^5$ and S. Anderson$^6$
           }
   \date{Received /  Accepted}

   \institute{$^{1}$Max-Planck-Institut f\"ur extraterrestrische Physik,
                   Postfach 1312, 85741 Garching, Germany\\
                   {$^2$}Institut d' Astrophysique de Paris, 98bis Bd Arago, F-75014 Paris, France \\
                   $^{3}$Johns Hopkins University, Department of Physics and Astronomy, Baltimore, USA \\
                   $^4$University of Maryland, Baltimore County,  1000 Hilltop Circle  Baltimore, MD  21250, USA \\
		   $^{5}$University of Crete, IESL, FORTH-hellas, 711 10, Heraklion, Crete, Greece \\
		  $^{6}$University of Washington, Dept. of Astronomy, BOX 351580, Seattle, WA 98195, USA  \\
  \\
}

   \offprints{Th. Boller (bol@mpe.mpg.de)}

   \date{Received /  Accepted}

   \maketitle
   \begin{abstract}

We present X-ray and optical observations of the X-ray bright
AGN 1ES 1927+654.  The X-ray observations obtained with ROSAT
and Chandra reveal persistent, rapid and large scale variations,
as well as steep 0.1-2.4 keV ($\rm \Gamma = 2.6 \pm 0.3$)
and 0.3-7.0 keV ($\rm \Gamma = 2.7 \pm 0.2$) spectra.
The measured intrinsic neutral X-ray column density is approximately
$\rm 7 \cdot 10^{20}\ cm^{-2}$. The X-ray timing properties indicate that the strong
variations originate from a  region, a few hundred light
seconds from the central black hole, typical for type 1 AGN.

High quality optical spectroscopy reveals a typical Seyfert 2
spectrum with some host galaxy contamination and no evidence of Fe II
multiplets or broad hydrogen Balmer wings.  The intrinsic optical
extinction derived from the BLR and NLR are $\rm A_V \ge 3.7$ and $\rm A_V=1.7$,
respectively.
The X-ray observations give an $\rm A_V$ value of less than
0.58, in contrast to the optical extinction values.

We discuss several ideas to explain this apparent difference in
classification including partial covering, an underluminous BLR
or a high dust to gas ratio.


\end{abstract}

\section{Introduction}

1ES 1927+654 was first detected as a X-ray source during the  Einstein
Slew Survey observations with a cumulative 38 s exposure and a count rate
of 0.93 $\rm counts$ s$^{-1}$
(Elvis et al. 1992).
Perlman et al. (1996) have  measured a redshift of 0.017 for 1ES 1927+654. They have also
demonstrated that the object was not a BL-Lac as previously classified, but a narrow-line radio galaxy. Bauer et al. (2000)
have confirmed this and further classified 1ES 1927+654 as a Seyfert 2.
The 6-cm radio flux density of 1ES 1927+654 is 16.03 mJy (Perlman et al. 1996) and the
optical B magnitude is 15.37.
Although the radio to optical flux density ratio is R = 3.3 (c.f. Kellermann et al. 1989),
the low radio power of $\rm 10^{22}$ W Hz$^{-1}$ at 5 GHz confirms its radio-quiet nature.
The probability of
a relativistic jet causing the extreme X-ray flux variations is
very unlikely.

In this paper we report on the X-ray spectra of 1ES 1927+654 obtained
with ROSAT and recently with Chandra, as well as on the results of an optical
follow-up observation in October 2001.
1ES 1927+654 was detected
as a X-ray bright source with extremely rapid and strong X-ray variability
in the ROSAT All-Sky Survey (Voges et al. 1999)  by a systematic analysis of the
X-ray timing properties of All-Sky Survey sources.
Additionally, a PSPC pointed observation,
during the final days of the ROSAT satellite in December 1998,
revealed a strong X-ray flare.

The detection of giant X-ray flares on short time scale of less than about
one day is of great importance for the study of the innermost regions of
active galactic nuclei. Models for explaining the presence of giant X-ray
flares include Doppler boosting effects of X-ray hot spots on the accretion
disk (Sunyaev 1973, Guilbert et al. 1983), X-ray flares above
the accretion disk (e.g Reynolds et al. 1999) or partial obscuration
of Compton thick matter close to the central black hole (Brandt et al. 1999,
Boller et al. 1997, Boller et al. 2002). Giant X-ray flares are therefore
thought to give independent evidence for the presence of relativistic motions close
to the central black hole, similar to the conclusions drawn from the
asymmetric Fe $\rm K\alpha$ line (Tanaka et al. 1995).

Persistent, strong and rapid variability seems to be a rare phenomenon among radio-quiet
AGN and has previously been detected only in  the  narrow-line Seyfert 1 galaxy
IRAS 13224--3809 (Boller et al. 1997).
The object shows amplitude X-ray variability of a factor greater than 10, on
time scales less than about 1 day.
The physical origin of giant and rapid X-ray flares is most probably different
from the X-ray variability occuring on much longer time scales of
years, or longer, in active and non-active galaxies.

A value of the Hubble constant of $H_0 \rm = 70\ km\ s^{-1}\ Mpc^{-1}$
and a cosmological deceleration parameter of $q_0 \rm = \frac{1}{2}$
have been adopted throughout.

\section{The X-ray properties of 1ES 1927+654}

\begin{figure}
  \psfig{file=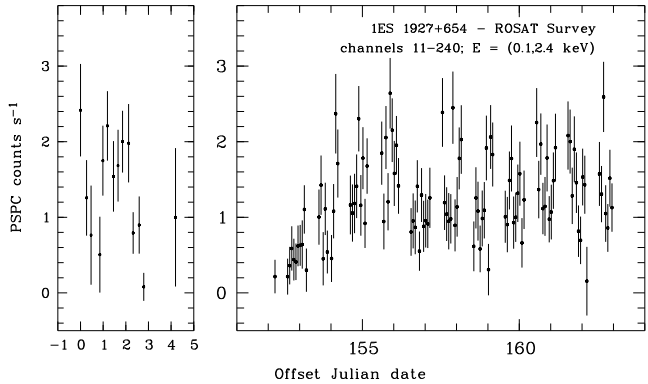,width=8.7cm,angle=-90,clip=}
  \caption{
ROSAT All-Sky Survey light curve of 1ES 1927+654 obtained
between 1990 July 11 (Julian date 2448084.1) and 1990 July 16
(left panel) and 1990 December 11 and 1990 December 21  (right panel).
Persistent, rapid and giant X-ray variability is detected. The maximum amplitude variation is a factor of 15, at
a significance level greater than 6 $\sigma$.
     }
\end{figure}

\begin{figure}
  \psfig{file=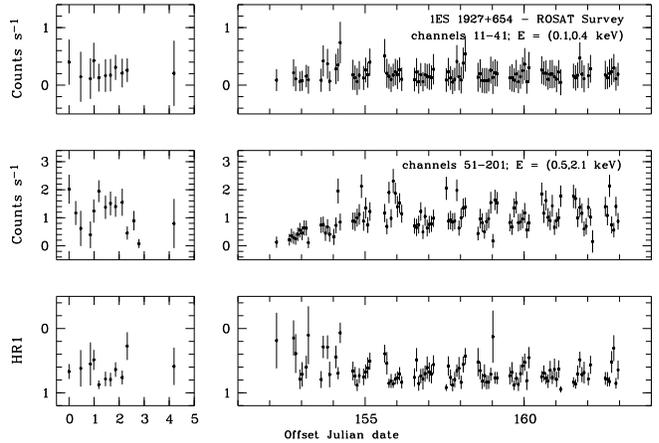,width=8.7cm,angle=-90,clip=}
  \caption{
  The two upper panels give the ROSAT All-Sky Survey
light curves
in the soft (channels 11--41) and hard (channels 51-201) ROSAT
energy bands referred to as S and H,
obtained between
between 1990 July 11 and 1990 July 16
and 1990 December 11 and 1990 December 2.
The lower panel gives the resulting hardness ratio HR1 = (H-S)/(H+S).
     }
\end{figure}

The X-ray properties of 1ES 1927+654 discussed below were obtained with the ROSAT and Chandra satellites.
The ROSAT data cover the survey observations carried out between July 1990 and July 1991 as well as a ROSAT
pointed observation obtained in December 1998.

For the ROSAT All-Sky Survey observations the source counts were obtained using a circular source cell with a radius
of 11.1 arcmin. The number of source plus background counts within this
cell was 4255 $\pm$ 65.
The background was determined from a source-free
cell with a radius of 11.5 arcmin, located in the scan direction through
the centroid position of the source and centered at
R.A.(2000) = $\rm 19^h22^m26.1^s$,
Dec.(2000) = $\rm 65^{\circ}44^{\prime}17^{\prime\prime}$.
The number of background counts normalized to the source cell size was
338 $\pm$ 18. The net counts are therefore 3917 $\pm$ 62, resulting
in a mean count rate of 1.22 $\rm \pm 0.02\ counts\ s^{-1}$.
For each path of the source through the ROSAT PSPC detector the
corresponding background was subtracted.

1ES 1927+654 was again observed during the final observation period of the
ROSAT satellite
in 1998 December 8 between 18:17:21 UT and 18:41:01 UT
with the PSPC detector. The total exposure time was 1338 seconds.
This observation was affected by some anomalies (cf. Sect. 2.1 of Boller et al. 2000) and we have performed
a careful data quality check to ensure that our results were not
affected.

In addition,
1ES 1927+654 was observed with Chandra during the guaranteed time programme.
The Chandra observation was performed using the
low energy transmission grating spectrograph (LETGS, see  Brinkman et al. 2000),
which comprises the low energy transmission grating (LETG)
and the high resolution camera microchannel plate detector
for spectroscopy (HRC-S).
1ES1927+654 was observed from March 20, 2001 at 13:20 UT
until March 21, 2001 at 06:58 UT.
The total exposure obtained from all good time intervals,
corrected for deadtime is 63110\,sec.
During the whole observation no times with high background
were detected.
The raw data was reprocessed and the spectrum extracted
using CIAO version 2.2.1 and CALDB release 2.15
following the corresponding science threads available at
the Chandra X-ray Center\footnote{\tt http://cxc.harvard.edu}.

The overlapping higher spectral orders were taken into account
by using the LETGS grating response matrices (version July, 2002)
which include 1st to 6th order.  Separate files are provided
for the negative and positive
orders\footnote{see \scriptsize http://cxc.harvard.edu/cal/Links/Letg/User/Hrc\_QE/\\
ea\_index.html}.
The fits with the same parameters were performed simultaneously
with the extracted negative and positive orders.

The Chandra X-ray centroid position of 1ES 1927+654 is $\rm R.A.(2000) = 19^h 27^m 19.61^s, Dec.(2000) = 65^{\circ}33^{\prime} 54.86^{\prime\prime}$,
which is in excellent agreement with the optical position (c.f. Fig. 9). 
The positional accuracy is about one arcsec.
The unprecedented positional accuracy of Chandra confirms the identification
of the strongly variable X-ray source with the distant galaxy.
All optically bright objects in Fig. 9 (labeled as 2-6) have been spectroscopically identified as late
G or early K stars. This rules
out any significant contribution of them to the X-ray flux of 1ES 1927+654.
The positional accuracy of ROSAT is less sensitive than Chandra,
however, the
centroid position of 1ES 1927+654 in the  ROSAT All-Sky Survey
of
R.A.(2000) = $\rm 19^h27^m19.2^s \pm 1.3^s$,
Dec.(2000) = $\rm 65^{\circ}33^{\prime}58^{\prime\prime} \pm 20^{\prime\prime}$
is consistent with the Chandra position.

\subsection{The strong, persistent and rapid X-ray variability of  1ES~1927+654}

In the following we constrain the timing properties obtained from the ROSAT All-Sky Survey,
ROSAT pointed and Chandra observations.
The plotted error bars in the light curves correspond to 1 $\rm \sigma$ in the Poisson regime (c.f. Gehrels 1986).
As a conservative approach we have calculated the total errors of  the counts using
the relation: $\rm 1.0 + \sqrt{(counts\ +\ 0.75)}$.

\begin{figure}
  \psfig{file=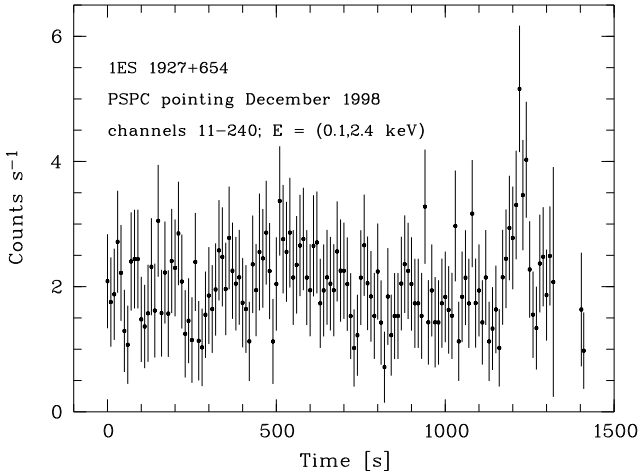,width=8.7cm,angle=-90,clip=}
  \caption{ROSAT PSPC pointed light curve obtained in
December 1998. A strong X-ray flare lasting for about 100 s
is detected near the end of the observations. The corresponding
isotropic energy is $\rm 10^{46}\ erg$ (see Sect. 2.1).
     }
\end{figure}

\begin{figure}
  \psfig{file=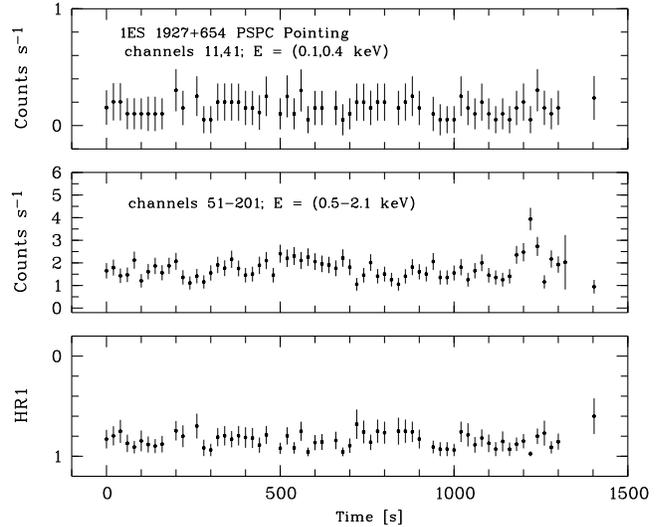,width=8.7cm,angle=-90,clip=}
  \caption{
ROSAT PSPC pointed  light curves
in the soft (channels 11--41) and hard (channels 51-201) ROSAT
energy bands
obtained on 1998 December 8.
The lowest panel gives the resulting HR1 hardness ratio.
}
\end{figure}

\begin{figure}
  \psfig{file=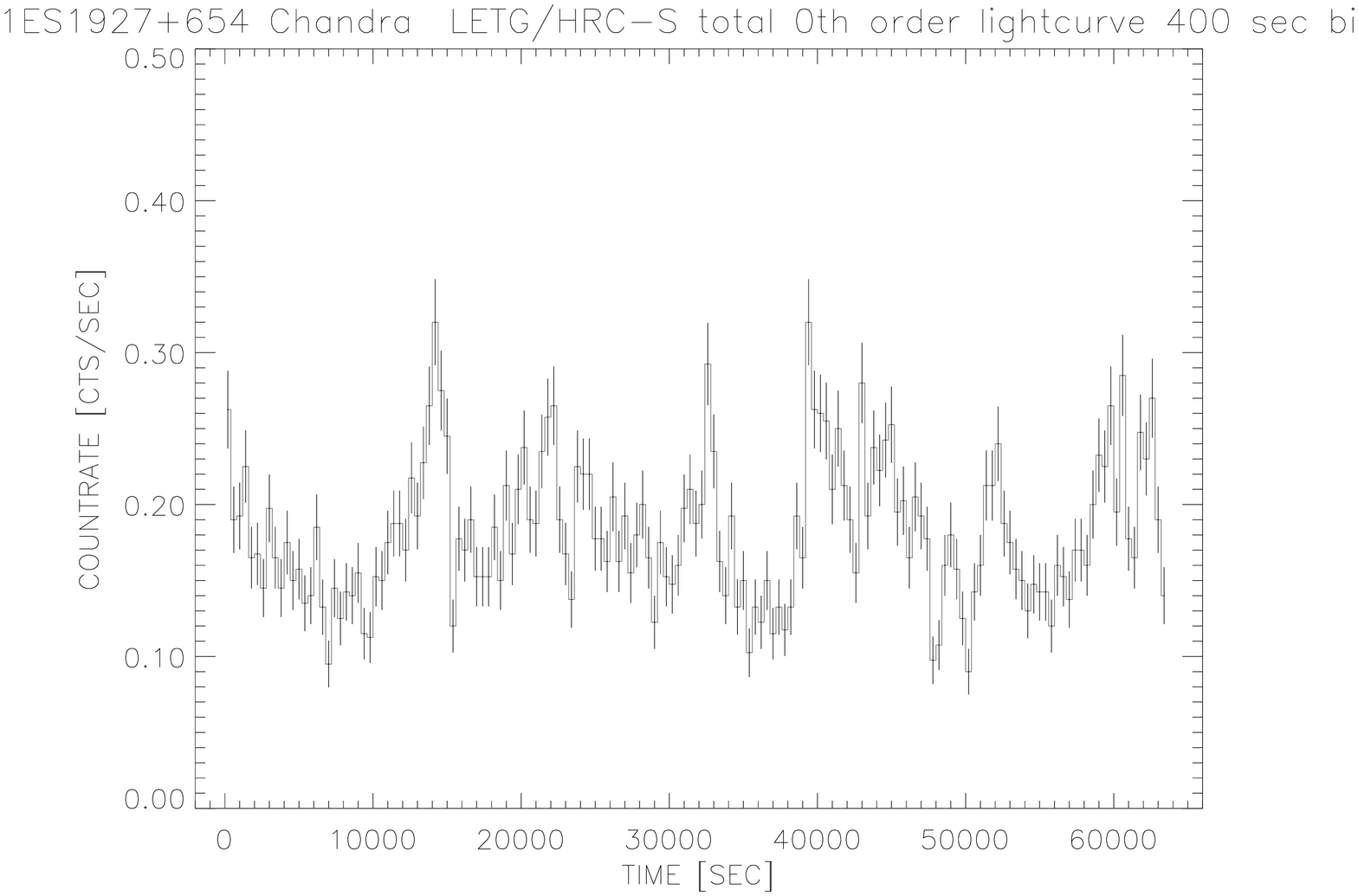,angle=90,width=8.7cm,clip=}
  \caption{The Chandra LETG light curve  in the 0.3--7 keV band
  also shows several strong flaring events with doubling time
  scales less than 400 s.
}
\end{figure}

1ES~1927+654 was observed during the ROSAT `mini-survey' for about 5 days
between 1990 July 11 and 16 with
a total exposure time of 254 seconds.
During the normal survey scan operations, 1ES~1927+654
was observed for about 11 days
between
1990 December 11 and 21.
Fig. 1 shows the \rosat PSPC survey light curve of 1ES~1927+654.
The left panel refers to the `mini-survey' observations in July 1990
and the right panel gives the count rate variations during the
survey scan observations in December 1990.
During the mini-survey the paths of 1ES 1927+654 through the PSPC
detector lasted only between 5.6 and 7.8 seconds and the source
passed the PSPC detector 49 times.
During the December observations
the source passed the PSPC detector 143 times and the exposure times
of the  individual scans range between 5.6 and 26.1 seconds.

Summing up all paths of 1ES 1927+654 through the PSPC detector during
all survey observations, results in a total exposure time of 3200 seconds.
Most interestingly are the unusually large amplitude and persistent variability,
making this object
the second radio-quiet
AGN showing this type of behavior, the first being
IRAS~13224--3809 (Boller et al. 1997).
As the exposure time per path during
the `mini-survey' is only about 1/3 of the
December 1990 observations, we have rebinned the light curve shown
in the left panel of Fig. 1 by summing up three paths of the source through
the PSPC detector, resulting in comparable exposure times per data point between
the two surveys. The mean separation between  the data points
is 4.8 hours in the left panel and 1.6 hours in the right panel.
Averaging the four data points with count rates below
0.3 $\rm counts\ s^{-1}$, gives a mean number of counts of
3.76 $\rm \pm$ 3.12, collected within 22.93 s. The four data points above
2.4 $\rm counts\ s^{-1}$ result in a mean number of counts of
60.18 $\rm \pm$ 8.80, obtained within  a 23.74 s exposure interval.
The resulting count rates are 0.164 $\rm \pm$ 0.136 and 2.534 $\rm \pm$ 0.370
$\rm counts\ s^{-1}$, for the low and  high states, respectively.
The maximum amplitude variability
is of a factor of 15, at  a significance level greater than 6~$\rm \sigma$.

In the following we constrain the spectral variability of 1ES 1927+654
based on the examination of appropriate hardness ratios.
In Fig. 2 we show the light curve of 1ES 1927+654 during the 1990--1991 All-Sky
Survey observations in the soft (channels 11--41; 0.1--0.4 keV) and hard (channels 51--201; 0.5--2.0 keV)
ROSAT energy bands, as well as the resulting hardness ratio light curve.
A constant model fit to the data points in the top panel
($\chi^2$ = 55.2 for 10 d.o.f)
and the middle panel of Fig. 2 ($\chi^2$ = 180 for 70 d.o.f)
can be rejected with $>$ 99.9 per cent confidence.
Significant spectral variability is detected during the
ROSAT All-Sky Survey observations of 1ES 1927+654.
The variations in the soft and hard ROSAT energy bands are not
correlated with the variations of the hardness ratio, making it
difficult to draw further conclusions of the underlying physical
emission mechanism causing the strong X-ray variability.
This is confirmed by performing simple power law fits to
different count rate intervals (e.g. from 0--1.5 and 1.5--3
$\rm counts\ s^{-1}$) and finding no significant differences in the
photon index. Simple relations as found 
in other variable AGN, e.g. the source becomes steeper when
the flux increases, do not seem to occur in 1ES 1927+654.

In Fig. 3 we present the ROSAT PSPC light curve on 1ES 1927+654.
Using a bin size of 10 seconds, a strong X-ray flare becomes apparent
between 1110 and 1280 seconds after the beginning of the observations.
The count rate fluctuations significantly exceed that of the variations
caused by the ROSAT wobble.
We have searched for associated spectral variability during the
flaring event. In Fig. 4 we demonstrate that the strong flux
variability is not correlated with significant spectral changes,
however, the hardness ratio light curve is still significantly
variable (a constant model fit can be rejected with $>$ 99.9 per cent
confidence).

The Chandra light curve (Fig. 5) indicates rapid count rate variations
with doubling time scales down to about 400 seconds.
The Chandra
timing properties further support the interpretation that the
variable X-ray emission arises within a few Schwarzschild radii
from the central black hole.

\subsection{Spectral properties}

\begin{figure}
  \psfig{file=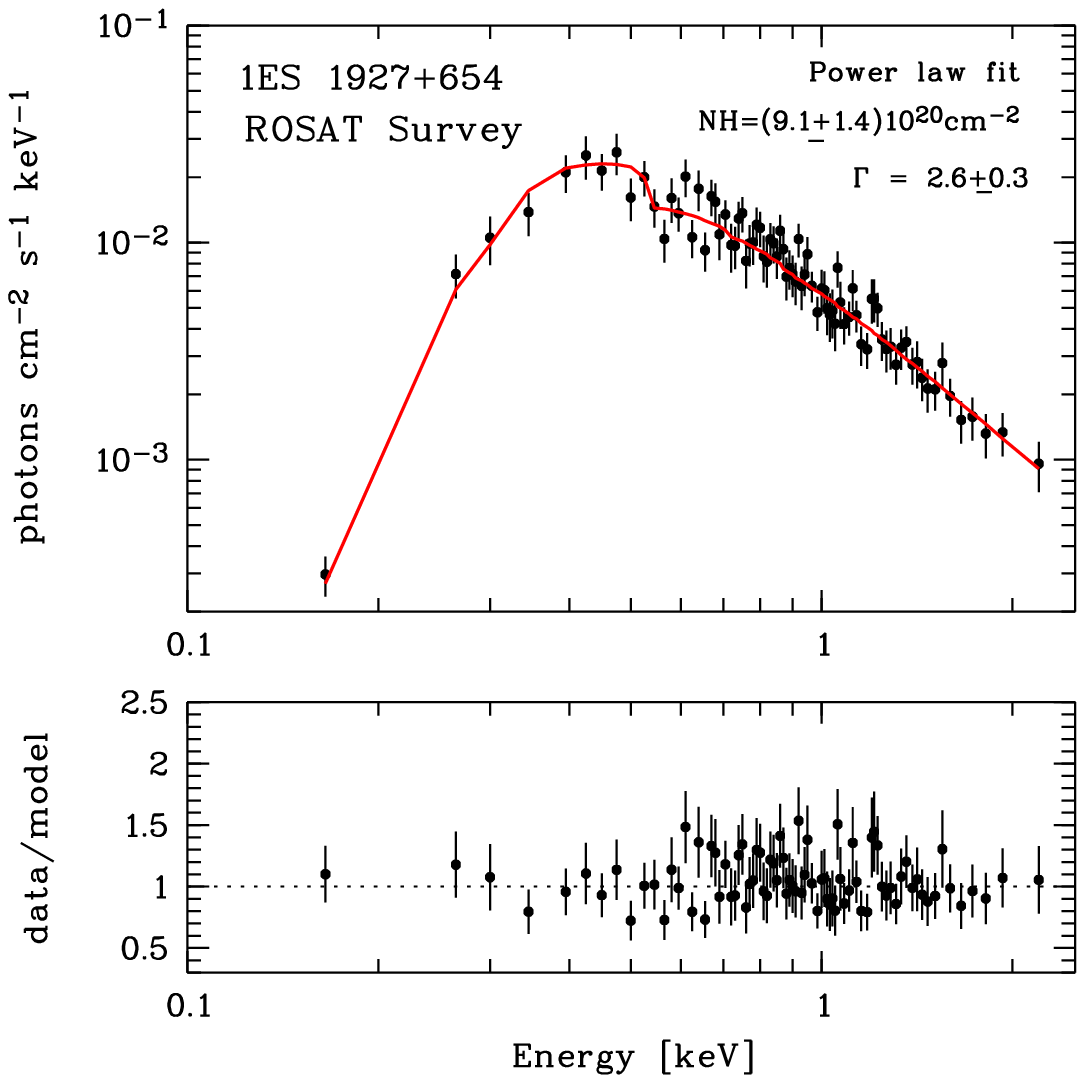,width=8.7cm,clip=}
  \psfig{file=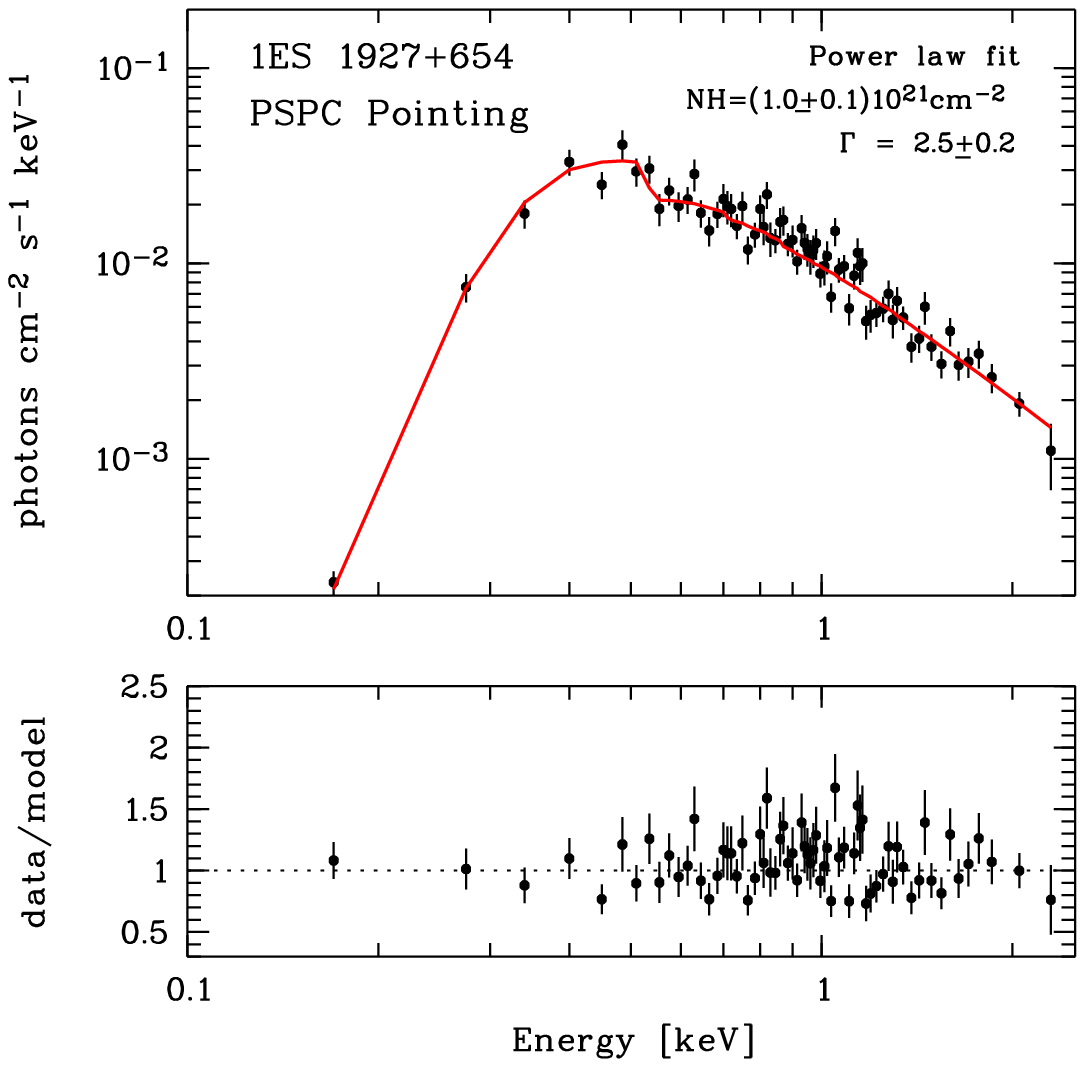,width=8.7cm,clip=}
  \caption{
Power-law fit to the ROSAT All-Sky Survey observations as well as to the pointed observations
with their corresponding
residua. The neutral absorbing column density is significantly above the
the
Galactic column density  of $N_{\rm H,Gal} \rm = (7.2 \pm 0.4) \cdot 10^{20}$ cm$^{-2}$
(Dickey \& Lockman 1990, Elvis et al. 1994).
}
\end{figure}

\begin{figure}
  \psfig{file=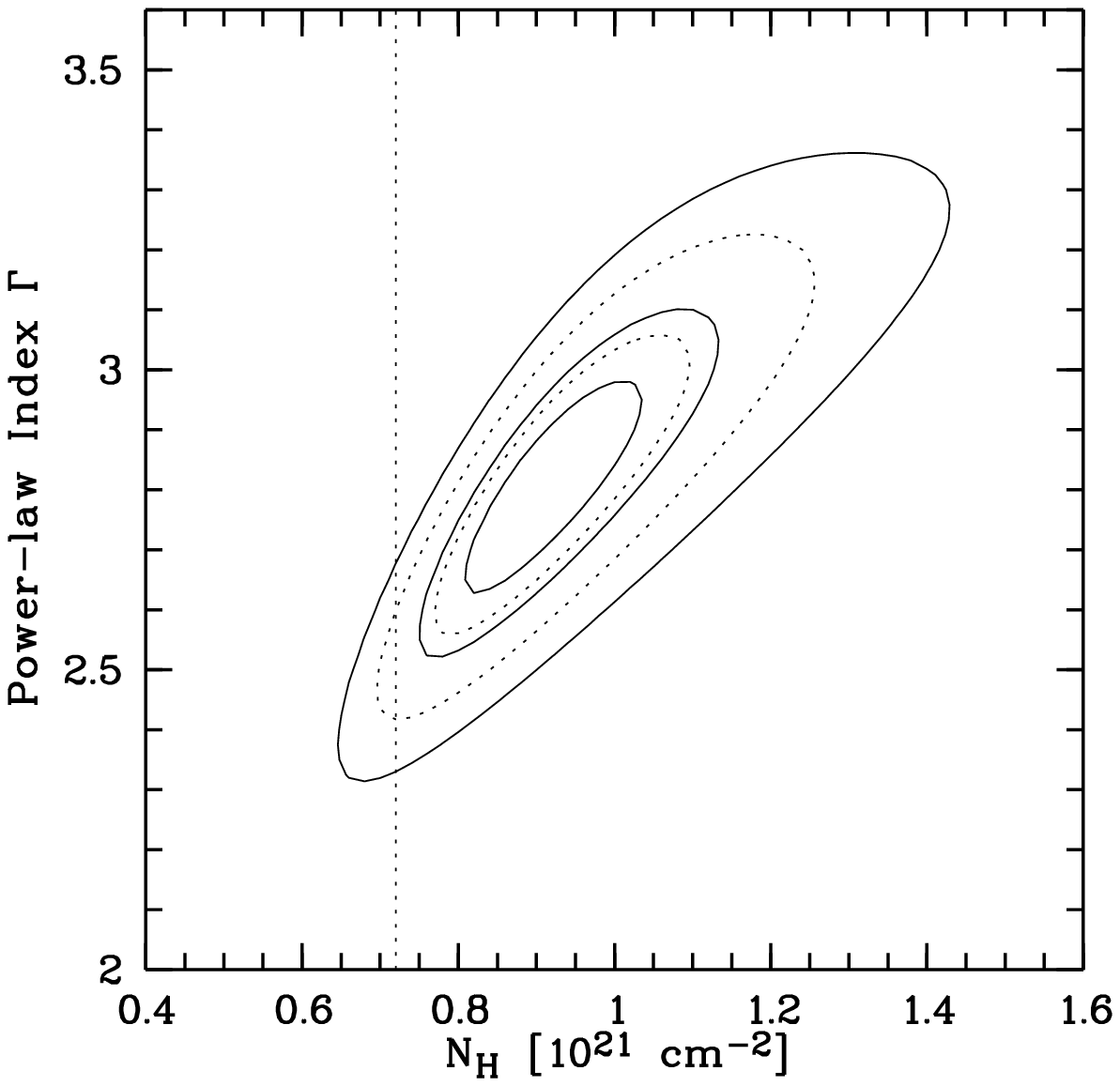,width=8.7cm,clip=}
  \psfig{file=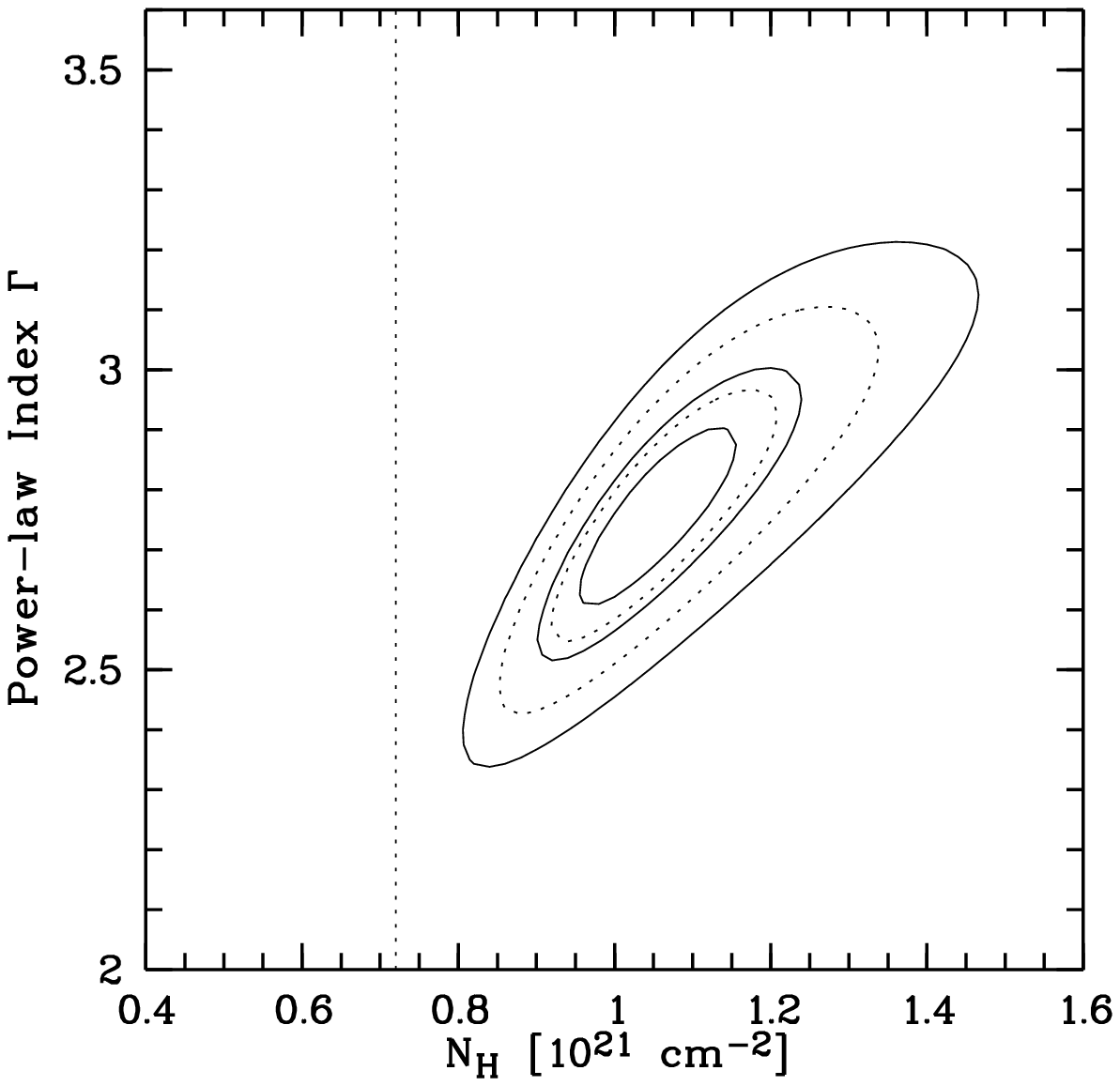,width=8.7cm,clip=}
  \caption{
Upper panel: Contour plot of $\chi^2$ as a function of the absorbing column
$N_{\rm H,fit}$ and the photon index $\Gamma$ for five confidence levels
of 68.3,95.4,99.7,99.99 and 99.99999 per cent for the ROSAT Survey observations
(upper panel) and the ROSAT PSPC pointed observations (lower panel).
The absorption is larger than the Galactic column (dashed vertical line)
at the 3 $\sigma$ level and at the 5 $\sigma$ level during the
pointed observations.
     }
\end{figure}

\begin{figure}[h]
  \psfig{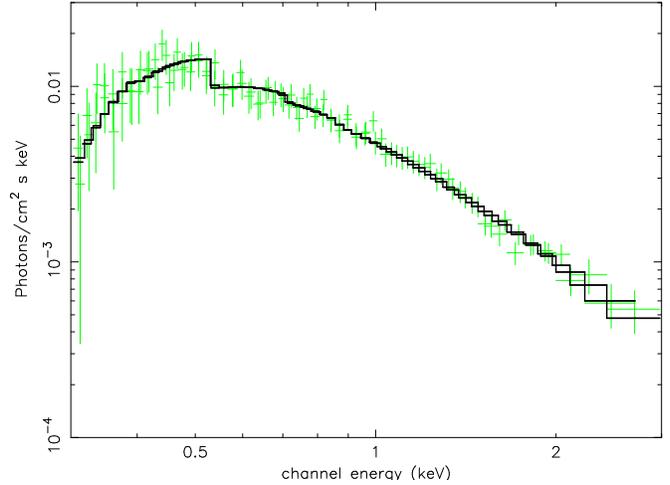}
  \caption{
  Power-law fit to the Chandra LETG observations of 1ES 1927+654. The spectrum was
  fitted by leaving the neutral oxygen abundance free. The oxygen abundance is found to be consistent with
  solar values.
  If we fix the Galactic absorption to $\rm 7.2 \cdot 10^{20}$ cm$^{-2}$, we find an intrinsic absorbing column
 density of $\rm (7.3 \pm 0.3)\cdot 10^{20}$ cm$^{-2} $.
}
\end{figure}

As described in the previous Section, significant spectral variations
are present, however, the count rate changes do not seem to be correlated
with the variations in the hardness ratio. The spectral fitting results
given below, represent the time-averaged spectra from each dataset
in the 0.1--2.4 keV energy band for the ROSAT observations and in the 0.3--7.0 keV
band for the Chandra LETG observations, respectively.

A simple power-law fit, where the absorption column density and the photon
index are allowed to be free parameters, provides an acceptable fit to the
ROSAT All-Sky Survey PSPC data ($\chi^2$ = 66 for 82 d.o.f; cf. Fig. 6).
The soft X-ray absorption of $N_{\rm H,fit} \rm = (9.1 \pm 1.4) \cdot 10^{20} cm^{-2}$    is
larger than the Galatic column towards 1ES 1927+654
of $N\rm_{H,gal} = (7.2 \pm 0.4) \cdot 10^{20} cm^{-2}$ (Dickey \& Lockman 1990;
Elvis et al. 1994)
at the 3 $\sigma$ level. The photon index is 2.6 $\pm$ 0.3.
These results are
robust to changes in the number of data points included in the fit.
Using the spectral parameters for 1ES 1927+654, as displayed in Fig. 6,
results in a
mean 0.1--2.4 keV flux of  $f \rm = 6.6 \cdot 10^{-11}$
erg cm$^{-2}$ s$^{-1}$, corresponding to an isotropic luminosity of
$L = \rm 4.6 \cdot 10^{43}$ erg s$^{-1}$.

A simple power-law fit to the PSPC pointed observation is shown in Fig. 6 (lower panel).
The absorbing column during the pointed observation is in excess of the
Galactic column above the 5 $\sigma$ limit (Fig. 7). No significant changes
in the photon index are detected, however, the mean count rate during the
pointed observation is 2.03 $\rm counts\ s^{-1}$, a factor of
2 larger than the ROSAT All-Sky Survey observations in 1990.
The isotropic {\bf 0.1--2.4 keV} energy emitted in the
strong X-ray flaring events (c.f. Fig. 3)  is $E \rm = 1.7 \cdot 10^{46}\ erg$.

The spectral energy distribution of 1ES 1927+654 as obtained with Chandra can be fitted with a
weak black body (accounting for the soft X-ray excess emission)
with kT = 0.01 keV and a power law with a  photon index of $\rm (2.5 \pm 0.2)$.
However, the black body temperature remains unconstrained.
A simple power-law model  also provides an acceptable fit.
If we fix the Galactic absorption to $\rm 7.2 \cdot 10^{20}$ cm$^{-2}$, we find an intrinsic absorbing column
density of $\rm (7.3 \pm 3.0)\cdot 10^{20}$ cm$^{-2} $, which is about a factor of 2 greater than that found
during the ROSAT observation (however, the difference is less than 2 $\rm \sigma$).
As the Chandra LETG observation gives the highest intrinsic soft X-ray absorption, we use that value
for comparison with the extinction values derived from the optical observations.
The photon index in the 0.3--7.0 keV band is $\rm 2.7 \pm 0.2$.
The isotropic luminosity in the 0.3 to 7.0 keV band is $\rm 2.1 \cdot 10^{43}\ erg\ s^{-1}$.

The strongest absorption edge from neutral oxygen at 0.537 keV is clearly visible in the LETG
spectrum. To constrain the oxygen abundance relative to hydrogen, we have fitted the spectrum
by leaving the oxygen abundance free. The spectral fit is still acceptable but does not significantly
improve the spectral fitting results. The oxygen abundance is found to be consistent with solar values.
Other element abundances can not
be constrained with the present statistics of the Chandra LETG observation.

If we assume that
(i) the empirical relation between interstellar X-ray absorption and optical extinction
    of $A_V = 4.5 \cdot 10^{-22} N_H$  (Gorenstein 1975) applies
    to 1ES 1927+654;
(ii) that the dust grains are optically thin to X-rays, i.e. that the amount of X-ray absorption
      by the dust grains is not much different from that by an equivalent mass of material in
      gaseous form;
(iii) that the X-ray and optical radiation travel through the same matter;
(iv) that the obscuration is not a strong function of time and
(v) a Galactic gas to dust ratio, then
the neutral hydrogen column densities derived from the X-ray spectra can be converted
into X-ray $\rm A_V$ values.
The maximum intrinsic X-ray  $A_V$ derived from the ROSAT and Chandra observations is 0.33 (following Gorenstein 1975) or
0.58 (following Predehl and Schmitt 1995).

\section{Optical spectral properties of 1ES 1927+654}
\begin{figure}
  \psfig{file=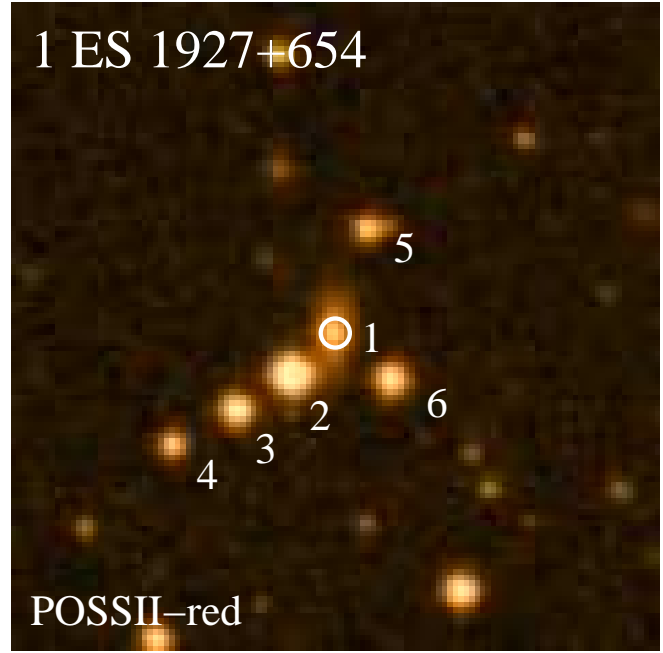,width=8.7cm,clip=}
  \caption{
POSS II red image of 1ES 1927+654. The 90 per cent error circle of the Chandra
position is overplotted. The field of view is 2 $\times$ 2 arcmin. The Chandra
position is within one arcsec consistent with the (1).
The objects
labeled with (2) to (6) are spectroscopically identified as G or K stars.
     }
\end{figure}

\begin{figure}
  \psfig{file=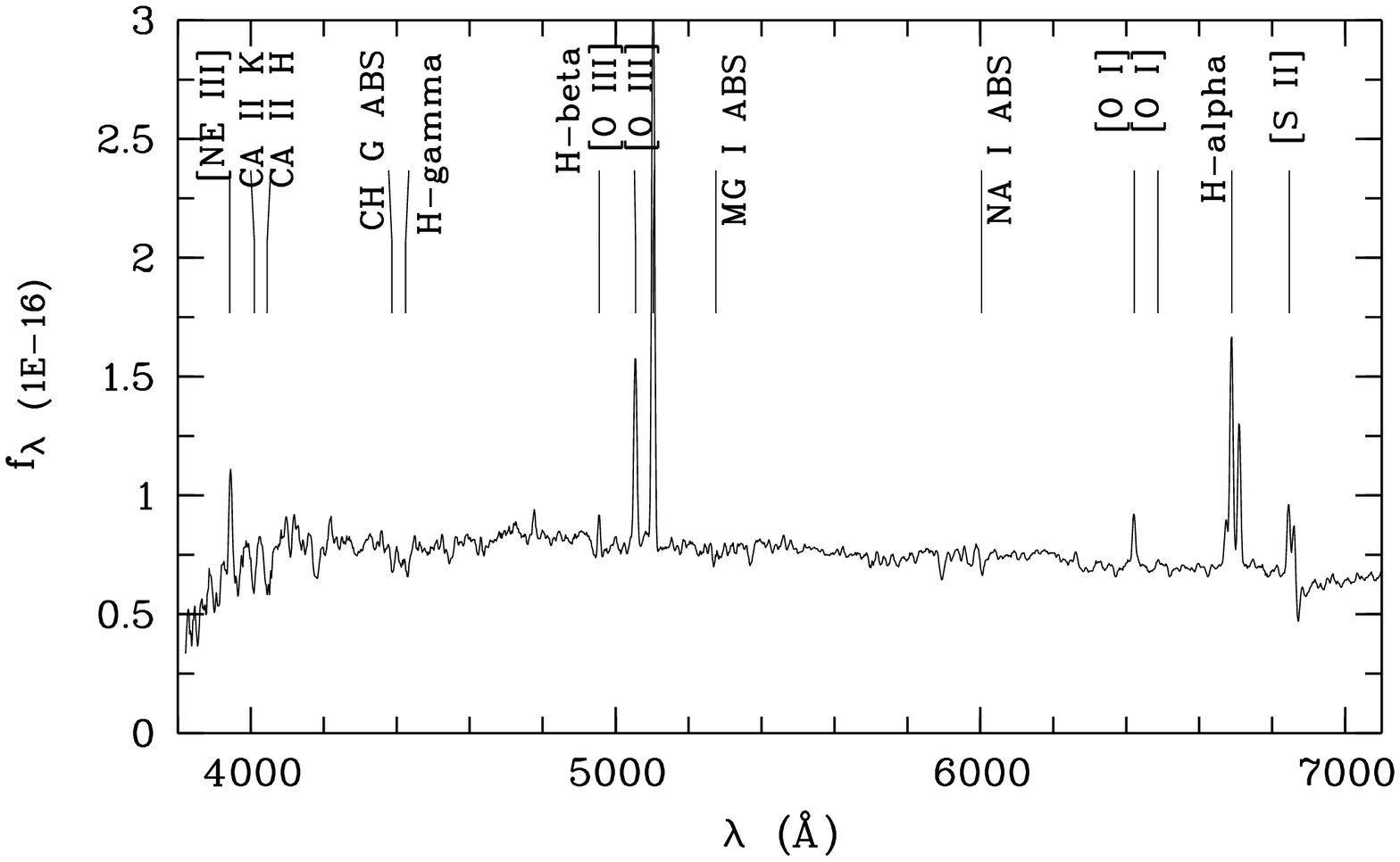,bbllx=40pt,bblly=70pt,bburx=405pt,bbury=655pt,width=8.7cm,angle=-90,clip=}
   \psfig{file=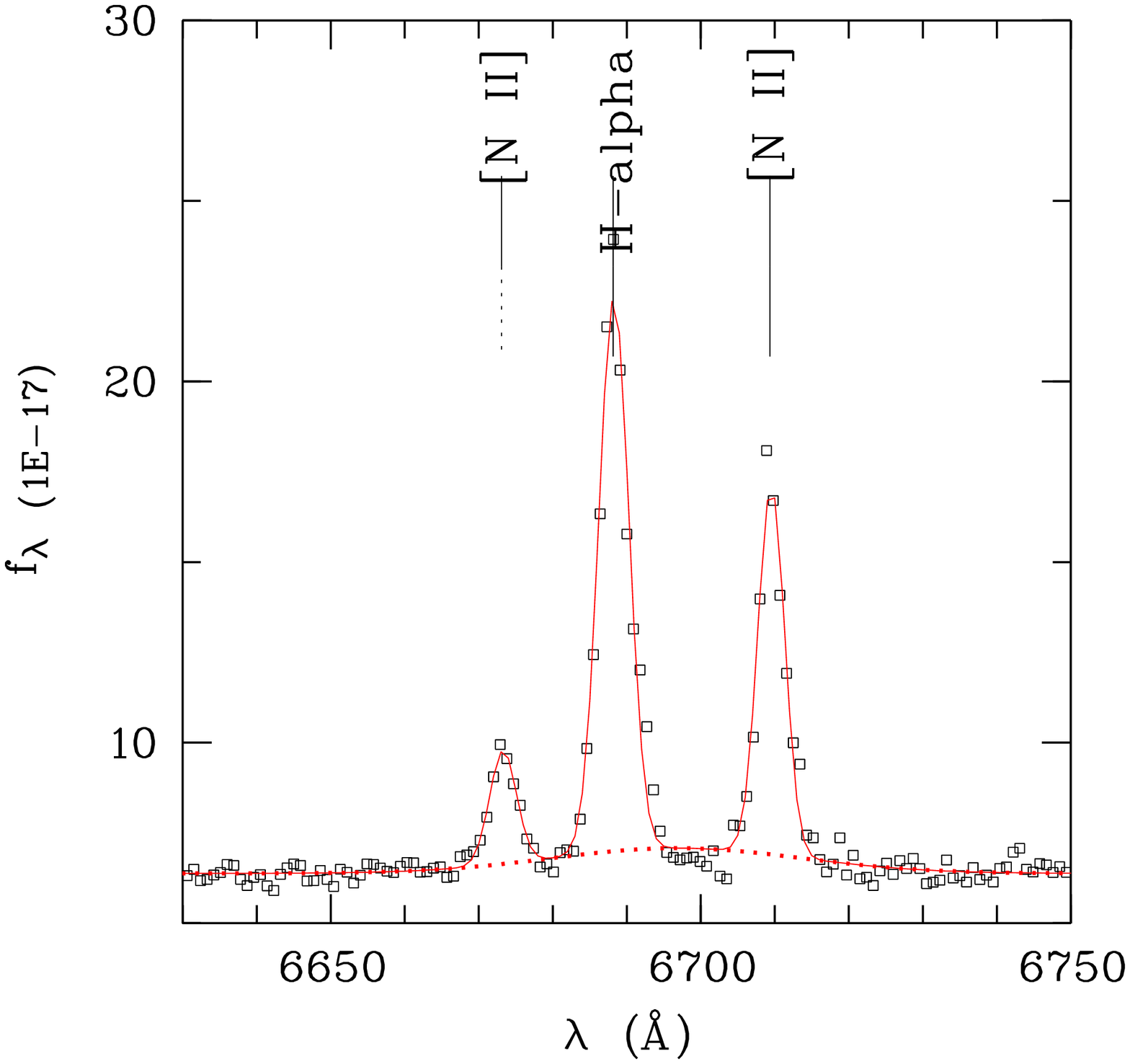,bbllx=40pt,bblly=30pt,bburx=545pt,bbury=590pt,width=8.7cm,angle=-90,clip=}
 \caption{Upper panel: Optical spectrum of 1 ES 1927+654 taken in June 2001.
Typical AGN (emission) and galaxy (absorption) lines are marked in the plot. The high [O III] $\lambda$5007 to
$\rm H\beta$ ratio, about 15, excludes a Narrow-Line Seyfert 1 classification.
Lower panel:
Spectrum of the H$\alpha$ region obtained in June 2002 with the same telescope and  instrument and
a spectral resolution of 3 \AA.
The H$\alpha$ region is well described by narrow emission lines (FWHM $\sim$ 170 km s$^{-1}$) plus a flat continuum
(solid line). A possible broad H$\alpha$ component with a maximum line flux of $\sim5$ \% of the narrow
H$\alpha$ emission line is overplotted (dotted line).
}
\end{figure}

The optical spectrum of 1ES 1927+654 was taken in June 2001
using the Carelec spectrograph with the 1.93m telescope at the Haute-Provence observatory.
The spectral resolution is 6 \AA~(3 pixels) and  the signal-to-noise ratio
is about 30.

The optical spectrum in Fig. 10 shows narrow emission lines
and several galaxy absorption line features of Ca H$+$K $\lambda\lambda3933/3968$,
CH G $\lambda4304$, Mg I $\lambda5173$ , and Na I $\lambda5893$ indicating the
presence of a host galaxy continuum.
The observed FWHM of most emission lines is about 330 km s$^{-1 }$, which
corresponds to the instrumental resolution.

The lower  FHWM value of $270\pm20$ km s$^{-1 }$ for the H$\beta$ line is probably due to the presence of an underlying absorption
component, which can barely
be seen when the spectrum is displayed at full resolution.
The flux ratios of [O III]5007/H$\beta=14.6$, [O I]6300/H$\alpha=0.2$,
[N II]6584/H$\alpha=0.6$, and [S II]/H$\alpha=0.6$ point to  a Seyfert 2 classification using
the standard diagnostics of Veilleux \& Osterbrock (1987).

The rest frame equivalent widths  of the absorption lines  CH G $\lambda4304$ and
Mg I $\lambda5173$ of $0.18\pm0.03$ \AA~and $0.14\pm0.03$~\AA, respectively, have been used to decompose
the spectrum into the
power-law featureless continuum and the host galaxy contribution. Assuming an optical
spectral index of $-1.5 \le \alpha_{opt} \le 2.0$ for AGN and  typical rest frame equivalent widths of
3.3 to 5.8 \AA~for CH G $\lambda4304$ and of 2.4 to 4.8 \AA~for Mg I $\lambda5173$
(see Goodrich \& Osterbrock 1983) we have
determined the fraction of the power-law featureless continuum to be between 92 and 97 \% at 4800 \AA.
The optical continuum emission of 1 ES 1927+654 seems therefore to be dominated by the AGN.

From the H$\alpha$ to H$\beta$ flux ratio we have determined the optical
extinction A$_V$=2.0 in the narrow line region, assuming Case B recombination (see Veilleux et al. 1997).
The actual visible extinction may in fact be even lower by a few percent, baring
in mind that the H$\beta$ emission line is affected by an underlying absorption,
which can not be precisely measured here. While the the optical spectrum does not clearly indicate the presence of a significant
broad H$\alpha$ component,
we have nevertheless tried to fit the H$\rm \alpha$ region with an additional component (see Fig. 10, lower panel),
to estimate its possible strength.
The addition of the broad component does not significantly improve the fit,
indicating that the presence of a broad $\rm H\alpha$
component is not required by the data. The line flux of any possible  broad component,
has to be lower than 5 per cent of the narrow H$\alpha$ component. The equivalent width ofthe  narrow  H$\alpha$ line
is 2.4 \AA.
 The central wavelength of this possible
broad H$\alpha$ emission component is also slightly shifted to the red.
We note that the lower signal-to-noise  spectrum of Perlman et al. (1996) also did not show a broad $\rm H_{\alpha}$ component.

Following  Goodrich et al. (1994),
the non-detection of a significant broad H$\alpha$ emission line suggests an extinction
value of the Broad Line Region (BLR) of at least $\rm A_V$ = 3.7.  Near-infrared spectra
of 1ES 1927+654 are required to estimate the upper limit of the extinction of the BLR by the detection
of broad emission line components at longer wavelength, e.g. at Br $\gamma$ or Br$\alpha$.

The $\rm A_V$ values discussed above refer to the sum of the Galactic and intrinsic extinction.
The Galactic $\rm A_V$ value in the direction of 1ES 1927+654 is 0.291 as found in NED. Therefore,
the corrected $\rm A_V$ for the Narrow Line Region in the distant AGN is at most 1.71 (or slightly lower if we could correct for the underlying absorption line at H$\beta$).

\section{Discussion}

The rapid, giant and  persistent  X-ray variability of 1ES 1927+654 detected with ROSAT and
confirmed by Chandra point to
a type 1 AGN classification, i.e. we have a direct view
to the central region. The optical spectra, however, give a
Seyfert 2 classification with some contribution from the host galaxy.
There seems to be a significant discrepancy (factor $\sim$ 6)
between the lower limit of  the $\rm A_V$ value
of 3.7 for the BLR, estimated from the optical spectrum, and
the maximum  $\rm A_V$ of 0.58 determined from
the X-ray spectrum.
We note that if  the dust is optically thin for X-rays the intrinsic X-ray  $\rm A_V$ can not greatly exceed the observed
value of 0.58, otherwise we would observe much stronger absorption edges as well as a higher energy soft X-ray cutoff.
In this section we will discuss the possible scenarios resulting in this unique combination of X-ray and optical properties.

\subsection{An underluminous BLR}
One possible explanation
for the apparent disagreement in X-ray and optical properties is an extremely underluminous, or even
absent, BLR.
The luminosity derived for the
putative broad
H$\alpha$ component (see Fig. 10) is about $\rm 10^{38} erg s^{-1}$, whereas
the total luminosity in the optical $B$-band is about $\rm 10^{43}\ erg\ s^{-1}$.  The typical ratio of
emission line to total B-band luminosity in Seyfert galaxies  is approximately 1 to 5 per cent (Netzer, private communication), thus
we would expect a line luminosity of about  $\rm 10^{41}\ erg\ s^{-1}$ in 1ES 1927+654.
This may be suggestive of an underluminous BLR.

\subsection {An optically thick X-ray absorber and/or higher dust to gas ratios}
The true optical extinction might be larger than that derived from the optical spectrum. It is possible
that, for example, the dust grains in the absorbing matter could be
individually thick to X-rays (cf. Fireman 1974). In this case, the soft X-ray spectrum cutoff does not
change with increasing $\rm A_V$ values, i.e. it remains constant in the X-ray observations. In addition, we could have
a large self-blanketing factor, in which case the absorbed luminosity from the BLR might
be even larger than the value derived in the case where the dust is optically thin to X-rays.
We also can not exclude that 1ES 1927+654 exhibits a higher dust to gas ratio than that of about 0.01 derived from
interstellar dust to gas ratios (Gorenstein 1975).

\subsection {Partial covering phenomena}
Partial covering phenomena have recently been detected with XMM-Newton in broad- and
narrow-line AGN (1H 0707-495, Boller et al. 2002; IRAS 13224-3809, Boller et al. in preparation;
PG 1211+654, Reeves, in preparation, and PDS 654, O'Brien, in preparation). Compton thick material partially
covers the central source, resulting in strong neutral or partially ionised Fe K absorption edges,
Fe K  re-emission features with fluorescent yields below  0.33, and strong soft X-ray emission.
However, the limited spectral coverage of the Chandra LETG spectrum does
not allow us to constrain partial coverer spectral components.
XMM-Newton is required to test this hypothesis. Most probably, the absorbing material has
to be Compton thick for X-ray and optical light (cf. Boller et al. 2002), which would not be in
disagreement with the optical extinction derived for 1ES 1927+654.

\subsection {Non-simultaneous X-ray and optical observations}
We note that the  X-ray observations were not taken simultaneously with the optical observations and
it is possible that the obscuration is a strong function of time. However, this possibility is not very likely as the X-ray observations are spread
over a time scale of 12 years which also
holds for the optical spectroscopic data (see e.g., Perlman et al. 1996). Whenever we observe 1ES 1927+654 at X-rays, it displays
strong and rapid variability. The optical spectrum also shows consistency over time.

\subsection {Future observations}
Near-infrared spectroscopy is required to further constrain the optical extinction for the BLR in 1ES 1927+654,
e.g. observations of Pa$\rm \beta$,  Pa$\rm \alpha$, Br$\rm \gamma$ and Br$\rm \alpha$.
Assuming that we will detect broad components in these lines, this will allow us to determine upper limits
for the $\rm A_V$ value up to about 68 (cf. Goodrich et al. 1994).

\vskip 0.5cm
We have presented an extreme and unique combination of optical and X-ray properties in 1ES 1927$+$654.
Clearly it is crucial to search for similar features in other galaxies. If such features are common in
other objects, this will further constrain the range of suitable models.

\begin{acknowledgements}
We would like to thank the anonymous referee for many helpful and constructive comments.
This research has made use of the NASA/IPAC Extragalactic Database (NED) which is operated by the Jet
             Propulsion Laboratory, California Institute of Technology, under contract with the National Aeronautics and Space
             Administration.
\end{acknowledgements}
%

\def\aa{A\&A}
\def\aas{A\&AS}
\def\aca{Acta Astron.}
\def\apj{ApJ}
\def\apjs{ApJS}
\def\asj{AJ}
\def\ibvs{Inf. Bull. of Var. Stars}
\def\mnras{MNRAS}
\def\ea{{et al.}}
\def\asp{Astron. Soc. of the Pac.}
\def\pasp{PASP}

{}
\end{document}